\documentclass[11pt]{article}
\usepackage{times}
\usepackage{geometry}
\usepackage[utf8]{inputenc}
\usepackage{enumitem,amssymb}
\usepackage{ragged2e}
\usepackage{graphicx}
\usepackage{comment}
\usepackage{multicol}
\usepackage[usenames]{xcolor} 
\usepackage{wrapfig}

\definecolor{xlinkcolor}{cmyk}{1,1,0,0}
\usepackage{url}
\usepackage[
 colorlinks=true,    
 linkcolor=xlinkcolor,     
 citecolor=xlinkcolor,     
 filecolor=xlinkcolor,  
 urlcolor=xlinkcolor,      
 final=true
]{hyperref}
\usepackage[super,sort&compress]{natbib}
\usepackage{enumitem}
\setenumerate{itemsep=0mm}

\begin{document}
\begin{raggedright} 
\huge
Snowmass2021 - Letter of Interest \hfill \\[+1em]
\textit{Hunting super-heavy dark matter with ultra-high energy photons} \hfill \\[+1em]
\end{raggedright}

\noindent {\large \bf Authors:}  \\[+1em]
Luis A. Anchordoqui (City University of New York, USA)\\
Corinne B\'erat (Grenoble Institute of Engineering Univ. Grenoble
Alpes, France)\\
Mario E. Bertaina (Universit\'a di Torino,  Italy)\\
Antonella Castellina (Osservatorio Astrofisico di Torino (INAF), Italy) \\
Olivier Deligny (Universit\'e Paris-Saclay, France) \\
Ralph Engel (Karlsruhe Institute of Technology,  Germany) \\
Glennys R. Farrar (New York University, USA)\\
Piera L. Ghia (Universit\'e Paris-Saclay, France) \\
Dan Hooper (Univresity of Chicago, USA)\\
Oleg Kalashev (Institute for Nuclear Research of the Russian Academy of Sciences,  Russia)\\
Mikhail Kuznetsov (Institute for Nuclear Research of the Russian Academy of Sciences, Russia)\\
Marcus Niechciol (Universit\"at Siegen, Germany )\\
Angela V. Olinto (University of Chicago, USA) \\
Philipp Papenbreer (Bergische Universit\"at Wuppertal, Germany)\\
Lorenzo Perrone (Universit\`a del Salento, Italy) \\ 
Julian Rautenberg (Bergische Universit\"at Wuppertal,  Germany)\\
Andr\'es Romero-Wolf (Jet Propulsion Laboratory, California Institute
of Technology, USA)\\
Pierpaolo Savina (Universit\`a del Salento, Italy) \\
Jorge F. Soriano (City University of New York, USA)   \\
Tonia M. Venters (NASA Goddard Space Flight Center, USA)

\noindent {\large \bf Abstract:} 

\noindent At any epoch, particle physics must be open to completely unexpected discoveries, and that is reason enough to extend the reach of searches for ultra-high energy (UHE) photons. The observation of a population of photons with energies  $E \gtrsim 100~{\rm EeV}$ would for example  imply the existence of either a completely new physical phenomena, or particle acceleration mechanisms heretofore never seen or imagined. But as we outline in this Letter of Interest, there are also good arguments for super-heavy dark matter (SHDM) in a parameter range such that it could be discovered via its decays to, in particular, UHE photons. Only ultra-high energy cosmic ray observatories have capabilities to detect UHE photons. We first investigate how current and future observations can probe and constrain SHDM models in important directions, and then outline some of the scenarios that motivate such searches. We also discuss connections between constraints on SHDM and on the parameter values of cosmological models.

\clearpage

At energies around 1 EeV ($10^{18}~$eV) and above, photons are expected to be  produced by $\pi^0$ decays, implying the existence of hadrons (that cause the production of $\pi^0$ mesons) with energies typically 10 times higher than the secondary photon. The search for such photons is of primary importance to decipher further the origin of UHE cosmic rays (UHECRs)~\cite{NeutPartLoI}, while the detection of photons of even higher energies, $E \gtrsim 100~{\rm EeV}$, would open an unexpected window, revealing either new physics or some new particle acceleration~\cite{Kotera:2011cp,Anchordoqui:2018qom,AlvesBatista:2019tlv}. As discussed in this Letter of Interest, the detection of a flux of UHE photons could be a smoking gun for dark matter (DM) composed of super-heavy particles. 

Multiple hypotheses have been proposed to describe DM, so far elusive. The leading benchmark relies on weakly-interactive massive particles (WIMPs) that were in thermal equilibrium in the early Universe~\cite{Bertone:2004pz}. The mass of these particles should lie in the range $10^2$-to-$10^4$ GeV so as to explain the DM density, which is inline with the naturalness argument to have new physics at the TeV scale. However, WIMPs have escaped any detection so far. The null results of direct detection push the originally expected  masses towards larger values and the couplings towards weaker ones. This gives increasingly strong constraints for the WIMPs to match the relic density. At the same time, so far, no new physics at the TeV scale have been observed at the LHC experiments~\cite{Rappoccio:2018qxp}. Although the exploration of the complete WIMP parameter space remains of great importance for the DM experimental program, a broader search program is also actively pursued. 

Models of super-heavy DM particles, first put forward in the 90s~\cite{Ellis:1990iu,Ellis:1990nb,Berezinsky:1997hy,Chung:1998zb,Kuzmin:1997jua,Birkel:1998nx,Berezinsky:1998ft}, were recently revived as an alternative to the WIMPs~\cite{Garny:2015sjg}. If super-heavy particles decay into standard-model particles, secondary products can be detected by CR observatories dedicated to UHECRs such as, currently, the Pierre Auger Observatory~\cite{ThePierreAuger:2015rma} and the Telescope Array~\cite{AbuZayyad:2012kk}, as well as next-generation experiments~\cite{Anchordoqui:2019omw,Romero-Wolf:2020vso,JHLOI}. Of particular interest would be the detection of UHE photons from regions of denser DM density such as the center of our Galaxy. Currently, the most stringent upper limits on photons around and above EeV energies come from the Pierre Auger Observatory, located near Malargue, Mendoza Province, Argentina.  

Although SHDM particles do not decay in a standard way because they are protected in the perturbative domain by the conservation of quantum numbers, they can disintegrate through non-perturbative effects. For non-commutative gauge theories, one of these effects can be the generation of one quantum number for the benefit of another through the change of configuration of gauge fields by tunnel effect (instantons) due to the vacuum structure~\cite{Kuzmin:1997jua}. At low temperatures, the probability of exciting topological field configurations from local charges can be estimated as $e^{-4\pi/\alpha_X}$, where
$\alpha_X$ is the coupling constant of the interaction considered at the natural scale of this interaction~\cite{tHooft:1976rip}. Therefore, if instantons are responsible for the decay of a particle, the lifetime of this particle scales as $\hbar e^{4\pi/\alpha_X}/m_Xc^2$. This  mechanism offers the possibility of providing metastable particles, which can produce detectable secondaries such as nucleons and photons. 
 
\begin{wrapfigure}{r}{0.5\textwidth}
    \includegraphics[clip, width=0.5\textwidth]{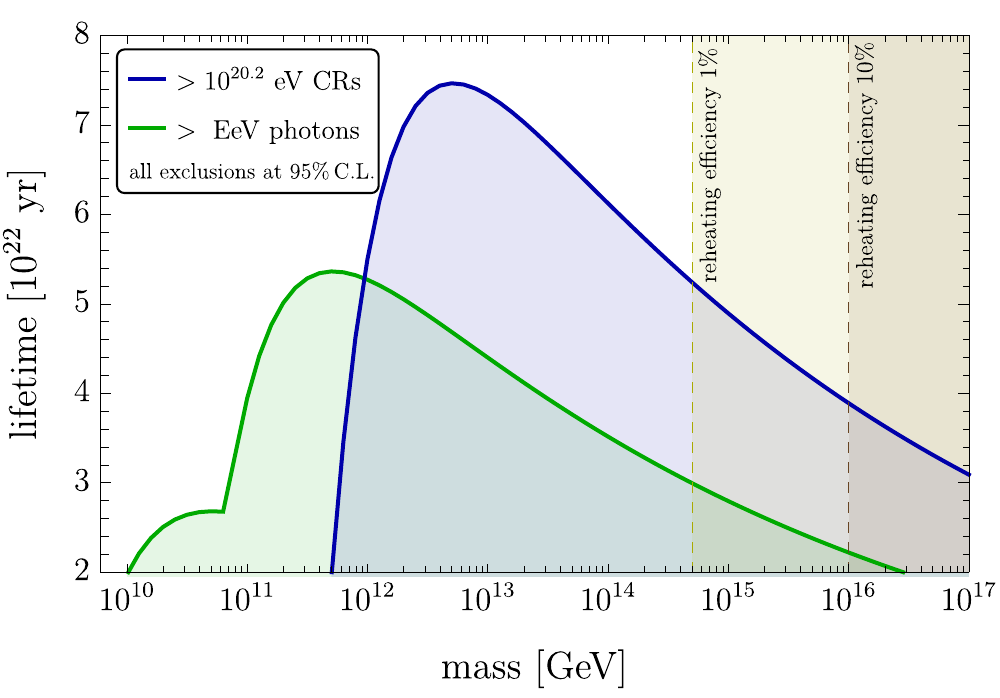}
    \caption{Constraints on the mass and lifetime of super-heavy DM particles from the absence of UHE photons (green) and from the absence of CR with energy above $10^{20.2}~$eV (blue). The allowed region lies above the curves. For illustration purpose, the 95\% CL upper limit on mass obtained from the possible value of the Hubble rate at the end of inflation for a reheating efficiency of 1\% (10\%) is shown as the vertical dashed (dotted) line~\cite{Garny:2015sjg}.}  
    \label{fig}
\end{wrapfigure}

No photons with energies above 1 EeV have been unambiguously identified so far~\cite{Aab:2016agp,Aab:2019ogu}. This can translate, see e.g.~\cite{Aloisio:2015lva,Kalashev:2016cre,Alcantara:2019sco}, into constrains on the properties of DM particles, as illustrated in Figure~\ref{fig} where the 95\% CL allowed regions of the mass and lifetime of the particles are shown. The green curve is obtained from upper limits on photons, while the blue one, more constraining at high masses, comes from the absence of UHECRs, hence also of photons, detected above $10^{20.2}$ at the Auger Observatory~\cite{Aab:2020rhr,Aab:2020gxe}. 

On the theoretical aspect, there are good motives for SHDM if new physics only manifest at the Planck scale or the GUT scale. This possibility is motivated not only by the absence of any sign of new physics at the TeV scale, but also by the precise measurements of the mass of the Higgs boson and of the Yukawa coupling of the top quark that make it possible to extrapolate the standard model all the way from the mass of the top to the Planck mass without encountering any inconsistency that would make the electroweak vacuum unstable. This vacuum lies in fact close to the boundary between stability and metastability~\cite{Degrassi:2012ry}. Alternatively to the naturalness solving the hierarchy question, the landscape of string theory vacua~\cite{Susskind:2003kw} could be at play to explain that several constants are abnormally small compared to the Planck scale. From anthropic aspects, some studies notably argue that the properties of nuclei and atoms would not allow the possibility of complex chemistry if the electroweak scale was too far from the confinement scale of QCD~\cite{Damour:2007uv}. In this case, there would no longer be any real reason for DM to be linked directly or indirectly to the electroweak scale. Although the  
structure formation constrains the DM density, it leaves a ``carte blanche'' for the mass spectrum of DM. The dark sector would be as natural as possible if the DM scale is related to the Planck scale or to the GUT scale.

SHDM particles that are only gravitationally coupled could have been produced at the end of inflation via the ``freeze-in mechanism''~\cite{McDonald:2001vt,Hall:2009bx,Bernal:2017kxu}, which relies on annihilations of the standard model particles to populate the dark sector. An interesting consequence is that, so as to produce enough such very feebly coupled heavy particles, the reheating temperature  must be relatively high, which implies a tensor/scalar ratio of the primordial modes possibly detectable in the power spectrum of the CMB. The limits inferred from the Planck satellite on this ratio thus constrain the possible phase space for the mass of the particles and the value of the Hubble rate at the end of inflation~\cite{Degrassi:2012ry}. The corresponding 95\% CL upper limits on the mass of SHDM, obtained from the Hubble rate at the end of inflation not to overshoot the CMB bounds on tensor modes, are shown as the vertical dashed and dotted lines in Figure~\ref{fig} for reheating efficiencies of 1\% and 10\%, respectively~\cite{Garny:2015sjg}. They are complementary to those obtained from the upper limits on UHE-photon fluxes. Conversely, the absence of  photons can be combined with cosmological models and data to constrain further the Hubble rate at the end of inflation as a function of the particle lifetime. 

Alternatively to the freeze-in mechanism to produce super-heavy DM particles, it is worth noting that a thermal freeze-out production could also be at play. Dynamical DM has been proposed, where different dark-matter components can interact and decay throughout the current epoch~\cite{Dienes:2011ja,Dienes:2011sa}. On the other hand, an annihilation rate that is exponentially enhanced relative to standard WIMPs could indeed be taking place if an additional hidden sector exists, through a co-annihilation with the lighter slightly-unstable hidden-sector species~\cite{Berlin:2017ife}. In this case, DM  decouples once the number density of the lighter species is sufficiently diluted by Hubble expansion, effectively delaying freeze-out. Then the search for UHE photons can also be used to constrain the parameter governing the decay of the lightest state in the hidden sector into visible-sector final states.

\textit{Summary.} It is now beyond doubt that accelerated particles by electromagnetic processes in astrophysical sites are responsible for the bulk of UHECRs. Yet a sub-dominant component could come from decay products of SHDM particles.  The continuous hunt for UHE photons with current and future UHECR detectors could thus lead to a serendipitous discovery of DM. The sensitivity to such a scenario is growing through, mainly, the bounds on UHE photons and the highest-energetic particles. The constraints are being more restrictive and the allowed parameter space is shrinking.


\clearpage

\begin{thebibliography}{99}

\bibitem{NeutPartLoI}
  Pierre Auger Collaboration,
{\it Fundamental Physics with Ultra-High-Energy Photons and Neutrinos at the Pierre Auger Observatory}, SNOWMASS21 - Letter of Interest (2020).

\bibitem{Kotera:2011cp}
K.~Kotera and A.~V.~Olinto,
Ann. Rev. Astron. Astrophys. \textbf{49}, 119-153 (2011)
doi:10.1146/annurev-astro-081710-102620
[arXiv:1101.4256 [astro-ph.HE]].


\bibitem{Anchordoqui:2018qom}
L.~A.~Anchordoqui,
Phys. Rept. \textbf{801}, 1-93 (2019)
doi:10.1016/j.physrep.2019.01.002
[arXiv:1807.09645 [astro-ph.HE]].



\bibitem{AlvesBatista:2019tlv}
R.~Alves Batista, J.~Biteau, M.~Bustamante, K.~Dolag, R.~Engel, K.~Fang, K.~H.~Kampert, D.~Kostunin, M.~Mostafa and K.~Murase, \textit{et al.}
Front. Astron. Space Sci. \textbf{6}, 23 (2019)
doi:10.3389/fspas.2019.00023
[arXiv:1903.06714 [astro-ph.HE]].
  
\bibitem{Bertone:2004pz}
G.~Bertone, D.~Hooper and J.~Silk,
Phys. Rept. \textbf{405}, 279-390 (2005)
doi:10.1016/j.physrep.2004.08.031
[arXiv:hep-ph/0404175 [hep-ph]].

\bibitem{Rappoccio:2018qxp}
S.~Rappoccio,
Rev. Phys. \textbf{4}, 100027 (2019)
doi:10.1016/j.revip.2018.100027
[arXiv:1810.10579 [hep-ex]].

\bibitem{Ellis:1990iu}
J.~R.~Ellis, J.~L.~Lopez and D.~V.~Nanopoulos,
Phys. Lett. B \textbf{247}, 257-264 (1990)
doi:10.1016/0370-2693(90)90893-B

\bibitem{Ellis:1990nb}
J.~R.~Ellis, G.~B.~Gelmini, J.~L.~Lopez, D.~V.~Nanopoulos and S.~Sarkar,
Nucl. Phys. B \textbf{373}, 399-437 (1992)
doi:10.1016/0550-3213(92)90438-H

\bibitem{Berezinsky:1997hy}
V.~Berezinsky, M.~Kachelriess and A.~Vilenkin,
Phys. Rev. Lett. \textbf{79}, 4302-4305 (1997)
doi:10.1103/PhysRevLett.79.4302
[arXiv:astro-ph/9708217 [astro-ph]].

\bibitem{Chung:1998zb}
D.~J.~H.~Chung, E.~W.~Kolb and A.~Riotto,
Phys. Rev. D \textbf{59}, 023501 (1998)
doi:10.1103/PhysRevD.59.023501
[arXiv:hep-ph/9802238 [hep-ph]].

\bibitem{Kuzmin:1997jua}
V.~A.~Kuzmin and V.~A.~Rubakov,
Phys. Atom. Nucl. \textbf{61}, 1028 (1998)
[arXiv:astro-ph/9709187 [astro-ph]].

\bibitem{Birkel:1998nx}
M.~Birkel and S.~Sarkar,
Astropart. Phys. \textbf{9}, 297-309 (1998)
doi:10.1016/S0927-6505(98)00028-0
[arXiv:hep-ph/9804285 [hep-ph]].

\bibitem{Berezinsky:1998ft}
V.~Berezinsky, P.~Blasi and A.~Vilenkin,
Phys. Rev. D \textbf{58}, 103515 (1998)
doi:10.1103/PhysRevD.58.103515
[arXiv:astro-ph/9803271 [astro-ph]].

\bibitem{Garny:2015sjg}
M.~Garny, M.~Sandora and M.~S.~Sloth,
Phys. Rev. Lett. \textbf{116}, no.10, 101302 (2016)
doi:10.1103/PhysRevLett.116.101302
[arXiv:1511.03278 [hep-ph]].

\bibitem{ThePierreAuger:2015rma}
A.~Aab \textit{et al.} [Pierre Auger],
Nucl. Instrum. Meth. A \textbf{798}, 172-213 (2015)
doi:10.1016/j.nima.2015.06.058
[arXiv:1502.01323 [astro-ph.IM]].

\bibitem{AbuZayyad:2012kk}
T.~Abu-Zayyad \textit{et al.} [Telescope Array],
Nucl. Instrum. Meth. A \textbf{689}, 87-97 (2013)
doi:10.1016/j.nima.2012.05.079
[arXiv:1201.4964 [astro-ph.IM]].

\bibitem{Anchordoqui:2019omw}
L.~A.~Anchordoqui, D.~R.~Bergman, M.~E.~Bertaina, F.~Fenu, J.~F.~Krizmanic, A.~Liberatore, A.~V.~Olinto, M.~H.~Reno, F.~Sarazin and K.~Shinozaki, \textit{et al.}
Phys. Rev. D \textbf{101}, no.2, 023012 (2020)
doi:10.1103/PhysRevD.101.023012
[arXiv:1907.03694 [astro-ph.HE]].

\bibitem{Romero-Wolf:2020vso}
A.~Romero-Wolf, J.~Alvarez-Mu\~niz, L.~A.~Anchordoqui, D.~Bergman, W.~Carvalho, A.~L.~Cummings, P.~Gorham, C.~J.~Handmer, N.~Harvey and J.~Krizmanic, \textit{et al.}
[arXiv:2008.11232 [astro-ph.HE]].

\bibitem{JHLOI}
J. Horandel {\it et al.}, {\it A next-generation cosmic-ray detector to study the physics and properties of the highest-energy particles in Nature}, SNOWMASS21 - Letter of Interest (2020).

\bibitem{tHooft:1976rip}
G.~'t Hooft,
Phys. Rev. Lett. \textbf{37}, 8-11 (1976)
doi:10.1103/PhysRevLett.37.8


\bibitem{Aab:2016agp}
A.~Aab \textit{et al.} [Pierre Auger],
JCAP \textbf{04}, 009 (2017)
[erratum: JCAP \textbf{09}, E02 (2020)]
doi:10.1088/1475-7516/2017/04/009
[arXiv:1612.01517 [astro-ph.HE]].

\bibitem{Aab:2019ogu}
A.~Aab \textit{et al.} [Pierre Auger],
[arXiv:1909.09073 [astro-ph.HE]].




\bibitem{Aloisio:2015lva}
R.~Aloisio, S.~Matarrese and A.~V.~Olinto,
JCAP \textbf{08}, 024 (2015)
doi:10.1088/1475-7516/2015/08/024
[arXiv:1504.01319 [astro-ph.HE]].

\bibitem{Kalashev:2016cre}
O.~K.~Kalashev and M.~Y.~Kuznetsov,
Phys. Rev. D \textbf{94}, no.6, 063535 (2016)
doi:10.1103/PhysRevD.94.063535
[arXiv:1606.07354 [astro-ph.HE]].

\bibitem{Alcantara:2019sco}
E.~Alcantara, L.~A.~Anchordoqui and J.~F.~Soriano,
Phys. Rev. D \textbf{99}, no.10, 103016 (2019)
doi:10.1103/PhysRevD.99.103016
[arXiv:1903.05429 [hep-ph]].

\bibitem{Aab:2020rhr}
A.~Aab \textit{et al.} [Pierre Auger],
Phys. Rev. Lett. \textbf{125}, no.12, 121106 (2020)
doi:10.1103/PhysRevLett.125.121106
[arXiv:2008.06488 [astro-ph.HE]].

\bibitem{Aab:2020gxe}
A.~Aab \textit{et al.} [Pierre Auger],
Phys. Rev. D \textbf{102}, no.6, 062005 (2020)
doi:10.1103/PhysRevD.102.062005
[arXiv:2008.06486 [astro-ph.HE]].

\bibitem{Degrassi:2012ry}
G.~Degrassi, S.~Di Vita, J.~Elias-Miro, J.~R.~Espinosa, G.~F.~Giudice, G.~Isidori and A.~Strumia,
JHEP \textbf{08}, 098 (2012)
doi:10.1007/JHEP08(2012)098
[arXiv:1205.6497 [hep-ph]].

\bibitem{Susskind:2003kw}
L.~Susskind,
[arXiv:hep-th/0302219 [hep-th]].

\bibitem{Damour:2007uv}
T.~Damour and J.~F.~Donoghue,
Phys. Rev. D \textbf{78}, 014014 (2008)
doi:10.1103/PhysRevD.78.014014
[arXiv:0712.2968 [hep-ph]].

\bibitem{McDonald:2001vt}
J.~McDonald,
Phys. Rev. Lett. \textbf{88}, 091304 (2002)
doi:10.1103/PhysRevLett.88.091304
[arXiv:hep-ph/0106249 [hep-ph]].

\bibitem{Hall:2009bx}
L.~J.~Hall, K.~Jedamzik, J.~March-Russell and S.~M.~West,
JHEP \textbf{03}, 080 (2010)
doi:10.1007/JHEP03(2010)080
[arXiv:0911.1120 [hep-ph]].

\bibitem{Bernal:2017kxu}
N.~Bernal, M.~Heikinheimo, T.~Tenkanen, K.~Tuominen and V.~Vaskonen,
Int. J. Mod. Phys. A \textbf{32}, no.27, 1730023 (2017)
doi:10.1142/S0217751X1730023X
[arXiv:1706.07442 [hep-ph]].

\bibitem{Dienes:2011ja}
K.~R.~Dienes and B.~Thomas,
Phys. Rev. D \textbf{85}, 083523 (2012)
doi:10.1103/PhysRevD.85.083523
[arXiv:1106.4546 [hep-ph]].

\bibitem{Dienes:2011sa}
K.~R.~Dienes and B.~Thomas,
Phys. Rev. D \textbf{85}, 083524 (2012)
doi:10.1103/PhysRevD.85.083524
[arXiv:1107.0721 [hep-ph]].


\bibitem{Berlin:2017ife}
A.~Berlin,
Phys. Rev. Lett. \textbf{119}, 121801 (2017)
doi:10.1103/PhysRevLett.119.121801
[arXiv:1704.08256 [hep-ph]].

\end{thebibliography}
\end{document}